\documentclass[12pt,dvips]{article}

\usepackage{amsmath,amssymb,exscale}
\usepackage{array,multicol}
\usepackage{afterpage,float,flafter}
\usepackage{epsfig,rotating,pifont}
\usepackage{cite}
\makeatletter
\@addtoreset{equation}{section}
\makeatother

\title{
\vspace*{-0.8cm}
\begin{flushright}
\end{flushright}
\vspace{1cm} 
\Large\textbf{GRAVITY'S \,\,SCALAR \,\,COUSIN}
\vspace*{.5cm}
\author{\large \textbf{
Raman Sundrum\footnote{email: ~ sundrum@pha.jhu.edu} 
}\\
\\
\emph{
Department of Physics and Astronomy} \\ 
\emph{Johns Hopkins University} \\ 
\emph{3400 North Charles St}. \\ 
\emph{Baltimore, MD 21218-2686}}}

\date{}
\begin{document}
\maketitle
\thispagestyle{empty}
\vspace*{.5cm}
  
\begin{abstract} 
The ``dilaton'', the Goldstone boson of spontaneously broken 
conformal field theories (in flat spacetime), is argued to 
provide a surprisingly provocative scalar analog of gravity. 
Many precise parallels and contrasts are drawn.
In particular, 
the Equivalence Principle, the Cosmological 
Constant Problem, and the tension between them is shown to be 
closely replicated. Also, there is a striking transition  
when mass is compressed within the (analog) 
Schwarzchild radius.
The scalar analogy may 
provide a simpler context in which to think about some of the puzzles
posed by real gravity.  
\end{abstract} 
  
\newpage 
\renewcommand{\thepage}{\arabic{page}} 
\setcounter{page}{1}


\section{Introduction}

A useful  analogy must  balance two requirements. It must obviously
share some key features with the  
original system of interest. But it should also 
exhibit significant 
differences, which are either simplifying in nature, or 
at least place the 
shared features in an illumninating new context. For example, this is what 
we would hope for if ever  we were to find  extra-terrestrial life. 
When the system of 
interest is gravity, in either the classical or quantum 
sphere, a worthy analogy may seem very hard to come by. 
General Relativity is simply too special. 
Nevertheless, it is the purpose of this paper to describe just such an analogy,
 in the hope that it will contribute 
to understanding the 
deep mysteries that underlie the real thing.

The analog system is simply the low-energy effective description of a
conformal field theory (CFT) in flat spacetime, whose conformal symmetry has 
somehow been spontaneously broken down to the usual Poincare symmetry. 
The low-energy ``chiral Lagrangian''
 describes the general couplings 
of the ``dilaton'', the scalar 
Goldstone boson of the symmetry breaking  (our 
analog of the graviton),  to other light remnants of the CFT 
(our analog of Standard Model matter). 
This paper will focus on 
pointing out the many surprising parallels between this scenario and 
real gravity. Important steps in this direction, including the development of 
the chiral Lagrangian, were taken long ago in 
Refs. \cite{salam} \cite{chiL} \cite{zumino} \cite{ellis}.
(Also see the discussions of scalar-to-gravity resemblance in Refs. 
\cite{dicke} and \cite{deser}.) Other ingredients appear in more 
recent literature.
The synthesis presented here is hopefully novel.
 The remainder of this introduction will simply
 summarize 
the parallels and contrasts to be drawn in the course of the paper.

$\bullet$ The dilaton is a Goldstone boson and real gravitons have also been 
previously viewed as Goldstone bosons \cite{chiL} \cite{per}. 
This particular parallel  
receives no further discussion in the rest of the paper.

$\bullet$ In the chiral Lagrangian description (Sections 2, 3)
there is an (emergent) fluctuating spacetime metric whose 
couplings to matter are constrained by general coordinate invariance 
\cite{chiL} \cite{zumino} \cite{ellis} \cite{rosen}. 
Consequently the Equivalence Principle is satisfied.  
There is a slight difference with real gravity (Sections 4, 5) 
in that there are small 
violations of the equivalence of gravitational and inertial masses due to the 
``gravitational'' self-energies of massive  bodies.

$\bullet$ Newton's Law holds in the regime of non-relativistic matter, 
constrained by the Equivalence Principle (Section 4).

$\bullet$ The theory is fully relativistic and there are relativistic 
corrections to Newton's Law but they differ in detail from those of 
General Relativity. There is dilaton radiation from accelerating matter 
but it differs from 
gravitational radiation in that it is scalar.

$\bullet$ When one tries to compress massive matter inside its Schwarzchild 
radius, something interesting happens (Section 5). 
In real gravity a black hole appears. 
In the dilaton theory the chiral Lagrangian description itself breaks down and
the full CFT dynamics becomes important. It is possible that 
a black-hole-like phenomenon still takes place. This is not proven but 
some evidence presented (Section 15).

 The holographic principle of real gravity \cite{holo} 
states that there are far fewer  degrees of freedom than naively appear in in our usual way of thinking about
 effective field theory. In the dilaton theory there is a sign that 
the naive number of degrees of freedom in the regime of validity of 
the  effective field 
theory is a gross overestimate, but this requires further study (Section 5).

$\bullet$ Light does not bend (classically) in a dilaton field (Section 6), 
unlike the 
classic effect in General Relativity. Therefore, there are no black holes 
in the conventional sense. Quantum effects can cause light to bend (Section 
7.1).

$\bullet$ The chiral Lagrangian description is a non-renormalizable 
effective quantum field theory which breaks down above the analog Planck 
scale just as effective General Relativity does (Section 7). 
The UV completion of the 
effective theory is a fundamental CFT with a moduli space (Section 11). 
For gravity the 
known UV completions are string theories. In both cases, supersymmetry 
appears in the known constructions. Remarkably, in some cases the two 
types of UV completions are related (Section 16) 
via the AdS/CFT correspondence \cite{adscft} (although 
the gravity is then higher dimensionsal).

$\bullet$ There is an analog of the cosmological constant which  
obstructs Poincare invariant solutions, but does give rise to 
solutions which are seen by matter to be de Sitter or anti-de-Sitter 
spacetimes (Section 8). 
Observers in the de Sitter solution will see Hawking radiation 
due to the cosmological horizon. 
Unlike real gravity, positive vacuum energy leads to the $AdS$ solution and 
negative vacuum energy leads to the $dS$ solution.

$\bullet$ General homogeneous and isotropic cosmologies can also 
be studied (Section 9) in the dilaton
chiral Lagrangian description. They are quite distinct from standard 
general relativistic cosmologies for simple reasons.

$\bullet$ There is a version of the Cosmological Constant Problem 
(Section 10), 
the matter contributions to the analog cosmological constant 
 being {\it identical} to the 
form of such contributions to the real cosmological constant 
(in the 
absence of subleading quantum dilaton and quantum gravity corrections).

$\bullet$ The analog Planck length acts as a minimal 
length scale (Section 12), certainly 
in the dilaton effective theory, and even in its UV completion. 

$\bullet$ Recently, effective field theory sense has been made of the notion 
of graviton mass \cite{gravmass} at the cost of exact general covariance, 
although the nature of any UV completion is unknown. In 
contrast there is a natural way in which a dilaton mass could arise 
(Section 13), but 
again at the cost of exact general covariance  in the couplings 
to matter.

$\bullet$ The supersymmetric version of the chiral Lagrangian description is 
straightforward to write down (Section 14) and is the analog of supergravity coupled 
to supersymmetric matter. Supersymmetry can technically protect the 
analog cosmological constant until supersymmetry is broken. There are 
interesting ways to do this.

$\bullet$ One can prove in the analog system that supersymmetry breaking 
implies the breaking of Poincare invariance \cite{susybreak} 
(Section 14), a property conjectured to be 
true in string theory \cite{banks}.

$\bullet$ The supergravity mechanism  of ``anomaly-mediation'' \cite{amsb} 
has rather close 
analogs in both the supersymmetric (Section 14) and non-supersymmetric 
(Section 7)
 effective descriptions of the dilaton coupled to quantum matter.

$\bullet$ Just as the Randall-Sundrum II model \cite{rs2} demonstrates the mechanism of 
localized 4D gravity in higher dimensions, there is a simple RS construction 
of a localized 4D dilaton (coupled to localized matter) in higher dimensions (Section 
15). In both 
cases the AdS/CFT correspondence illuminates the localization mechanism (Section 16).

$\bullet$ While attempts to define Euclidean quantum effective 
gravity are troubled by
an IR problem, namely the unboundedness of the Euclidean Einstein action 
\cite{euclgr}, 
the Euclideanized theory based on the dilaton chiral Lagrangian is 
well defined in the IR (Section 17). 
Of course both real and analog gravity suffer in the 
UV from non-renormalizability.

~

Effective field theories of the dilaton 
coupled to matter are sufficiently rich that they can mimic our real universe 
 in its everyday running. There is no difficulty at the 
level of the chiral Lagrangian in describing fields and interactions that 
support  stars, solar systems and life. 
Indeed, at the dawn of relativity, 
scalar theories were seriously 
 considered as realistic candidates for gravitational dynamics 
\cite{scalgr}.

\section{Dilaton Chiral Lagrangian}

Here, and in the next section, we will develop
the chiral Lagrangian  corresponding to 
the 
spontaneous breakdown of conformal symmetry to standard Poincare 
symmetry (in Minkowski spacetime). 
For the most part, this is
 a review of Refs. \cite{chiL} \cite{zumino} \cite{ellis}, 
but given in the most suggestive language for our purpose.
(In general, there are many possible, physically equivalent, formalisms for 
 chiral Lagrangians, connected by field redefinitions 
\cite{ccwz}.)

The fifteen parameter conformal group is  the subgroup of general coordinate 
transformations that
result in an overall coordinate-dependent rescaling of the Minkowski metric:
\begin{equation}
\label{conf}
ds^2 = \eta_{\mu \nu} dx^{\mu} dx^{\nu} =_{x'(x)}  f(x') 
\eta_{\mu \nu} dx'^{\mu} dx'^{\nu}. 
\end{equation}
It consists of 
 the Poincare transformations, rigid 
scale transformations and the special conformal transformations, 
\begin{equation}
x'^{\mu} \equiv \frac{x^{\mu} + x^2 b^{\mu}}{1 + 2 b.x + b^2 x^2}.
\end{equation}
It is generally believed that any unitary theory possessing Poincare and 
scale symmetry will automatically be invariant under the special conformal 
transformations as well. For a discussion see Ref. 
\cite{joe}. Relatedly, 
a single scalar Goldstone field for broken scale invariance, the ``dilaton'', 
 is all that is needed in order 
to non-linearly realize conformal invariance in the chiral lagrangian. The 
extra four Goldstone bosons corresponding to broken special conformal 
generators can be taken as the derivatives of the dilaton
  rather than independent 
fields \cite{chiL}. The breakdown of the one-to-one correspondence between Goldstone 
fields and broken generators can occur when these generators are 
constructed from fewer conserved 
currents, as can happen with spacetime symmetries \cite{low}.

 We will 
employ a dimensionless interpolating field for the dilaton,
\begin{equation}
\phi(x) \equiv e^{\pi(x)/M},
\end{equation}
where $M$ is the order parameter scale of conformal invariance breaking, 
and $\phi(x) M$ describes Goldstone fluctuations about this vacuum, with 
canonical field $\pi(x)$.
We clearly are 
expanding about the VEV $\langle \phi \rangle = 1$, any other constant choice
being absorbable into $M$. 
$\phi$ (or $\pi$) will be our analog scalar graviton and 
$M$ will play an 
analogous role to the Planck scale in real gravity. The quantitative value 
 of the scale $M$ is 
arbitrary by the fundamental scale invariance of the dynamics but for the 
sake of our analogy we will take it to be the same as the real Planck scale. 

The non-linear Goldstone
transformation law under conformal symmetry is conveniently taken to be 
\begin{equation}
\label{diltransf}
\phi'(x') = \sqrt{f(x')} \phi(x),
\end{equation}
where $f$ is given in Eq. (\ref{conf}). Thus for rigid scale transformations
 for example, $x' = \lambda x$,
\begin{equation}
\pi'(\lambda x) = \pi(x) - M \ln(\lambda).
\end{equation}
This is similar to the usual shift transformations of spontaneously broken 
internal symmetries, except that the shift in the field is accompanied by
 a change of the  coordinate on the left-hand side. This last point is 
 important when we write the most general chiral lagrangian for 
the dilaton because, unlike Goldstone bosons of internal symmetries, we 
are now able to write a (unique form of) dilaton potential, 
\begin{equation}
\label{dilS}
S_{dilaton} = \int d^4x \{
\frac{M^2}{2} (\partial_{\mu} \phi)^2 - \Lambda \phi^4 +
~ ~ {\rm higher ~ derivatives} \}.
\end{equation}
The conformal invariance is  straightforward to check. 

The appearance of the non-derivative coupling is unusual. Normally when 
we contemplate a symmetry of the dynamics, 
it seems a  robust possibility 
that it is realized in a spontaneously broken phase. Then we can write a 
chiral lagrangian for the requisite Goldstone field and its derivative 
couplings ensure that the Goldstone field manifold describes degenerate 
vacua. In particular any choice of Goldstone VEV is a suitable vacuum choice.
In the present case with the $\phi^4$ coupling, this is false, and in fact 
it results in a runaway behavior, $\phi \rightarrow \infty, 0$, where the 
conformal invariance
 is either broken at infinitely high energy or not broken at all. In 
either case we are driven out of the useful regime for a chiral lagrangian.
However, it may be that $\Lambda = 0$ ($\Lambda \ll M^4$), so that 
our vacuum choice,  $\langle \phi \rangle = 1$, is (approximately) 
stable. Very naively, we would have expected $\Lambda \sim {\cal O}(M^4)$ in the 
chiral lagrangian, which  shows us that the broken phase of conformal 
invariance requires special circumstances, not as robust as with other 
symmetries.  We shall nevertheless
proceed by assuming that $\Lambda \ll M^4$. 

A first connection with standard General Relativity arises by constructing a 
fluctuating ``auxiliary'' (as opposed to independent)
 metric out of the dilaton, 
\begin{equation}
\label{aux}
g_{\mu \nu}(x) \equiv \phi^2(x) \eta_{\mu \nu}, 
\end{equation}
and noting that the standard metric transformation law and the dilaton 
transformation precisely agree under conformal coordinate transformations. 
In terms of the auxiliary metric we can re-write the chiral lagrangian, 
\begin{equation}
\label{fakegr}
S_{dilaton} = \int d^4 x \sqrt{-g}\{- \frac{M^2}{12} R - \Lambda +
~ ~ {\rm higher ~ derivatives}\}.
\end{equation}
Of course it is understood that in varying this action  
$\phi(x)$ is the independent variable, not $g_{\mu \nu}$.
 In fact, 
the curvature (kinetic) term above has the opposite sign from the standard 
Einstein action. The latter sign in General Relativity 
gives positive energies to physical 
 gravitational fluctuations, while fluctuations of the  form 
$g_{\mu \nu}(x) \equiv \phi^2(x) \eta_{\mu \nu}$ appear to have 
negative-definite gradient energies, but fortunately these flucations 
 are pure gauge.
However in 
the dilaton theory, the dilaton is physical and there are no other 
fluctuations, and bounded energy is achieved by a sign flip.

Despite
the  general covariance  of the action, Eq. (\ref{fakegr}), general 
covariance is not respected by the condition, Eq. (\ref{aux}). However we can 
reformulate the chiral lagrangian in a completely generally covariant fashion
by taking $g_{\mu \nu}(x)$ to be the fundamental field, and imposing the 
generally covariant subsidiary condition that the Weyl tensor vanishes,
\begin{equation}
R_{\lambda \mu \nu \kappa} - \frac{1}{2} (g_{\lambda \nu} R_{\mu \kappa}
- g_{\lambda \kappa} R_{\mu \nu} - g_{\mu \nu} R_{\lambda \kappa} + 
g_{\mu \kappa} R_{\lambda \nu}) + \frac{R}{6} (g_{\lambda \nu} g_{\mu \kappa}
- g_{\lambda \kappa} g_{\mu \nu}) = 0.
\end{equation}
Metrics satisfying this condition  are precisely those which can 
be expressed in the form, Eq. (\ref{aux}), in some coordinate system 
\cite{weyl}.
Thus if we extremize the action for metrics satisfying this 
condition, we will recover the classical content of the dilaton theory.
At the level of quantum effective field theory
 we could impose the vanishing of the Weyl tensor in a path 
integral over metrics, which would be equivalent to integrating over 
$\phi$ along with some coordinate gauge redundancy, which can be gauge fixed 
in the usual way.  For the rest of this paper however,
it is sufficient to think of $\phi$ as the 
fundamental field and $g_{\mu \nu}$ as an auxiliary construct, useful in 
adding other light matter ($\ll M$) of the broken conformal theory to the 
chiral lagrangian description, as we will now see.

\section{Matter and the Equivalence Principle}

Once conformal invariance is spontaneously broken, one expects $M$ to set 
the mass gap for generic states made from the conformal theory. 
However there may be 
states which are much lighter than $M$, 
protected by infrared symmetries or 
by coincidence. We must therefore include this light matter sector
 in our 
chiral lagrangian description below $M$, coupling it to the dilaton in 
the most general conformally invariant way. As usual, weakly coupled
matter can only be
scalars $\chi$, fermions $\psi$, and vector fields $A_{\mu}$. 
 This matter will be 
our analog of Standard Model matter in the real world.
We can write 
the general\footnote{Without loss of generality, we have written the chiral lagrangian in 
``Einstein frame'' because the Weyl transformations of the metric needed 
to go to this frame 
obviously correspond to well-defined transformations on the dilaton by 
Eq. (\ref{aux}).}  conformally invariant low-energy effective action in the 
(notationally-condensed) form, 
\begin{eqnarray}
\label{full}
S_{eff} &=& \int d^4 x \sqrt{-g} \{ - \frac{M^2}{12} R + 
k_1(\chi, \psi) g^{\mu \nu} D_{\mu} \chi^*  D_{\nu} \chi +
k_2(\chi, \psi) \overline{\psi} i e^{\mu}_{~ a} D_\mu \gamma^a \psi   
- V(\chi, \psi) \nonumber \\
&-& k_3(\chi, \psi) g^{\mu \alpha} 
g^{\nu \beta} F_{\mu \nu} F_{\alpha \beta} 
+ i k_4(\chi, \psi) \frac{\epsilon^{\mu \nu \alpha \beta}}{\sqrt{-g}} 
F_{\mu \nu} F_{\alpha \beta}
- k_5(\chi, \psi) \overline{\psi} \sigma^{ab} \psi F_{\mu \nu}  e^{\mu}_{~ a}
e^{\nu}_{~ b} \nonumber \\
&+& k_6(\chi, \psi) e^{\mu}_{~ a} D_{\mu} \chi \overline{\psi} \gamma^{a} \psi
+ ~ {\rm higher ~ derivatives} \}.
\end{eqnarray}
The explicitly written terms are up to two-derivative order (one-derivative 
order when there are fermions present). 
The covariant derivative for fermions hides a spin connection 
based on the auxiliary vierbein, $e_{\mu}^{~ a} \equiv \phi(x) 
\delta_{\mu}^{~ a}$, but the scalar (gauge-covariant) derivatives and gauge 
field strengths are independent of the metric (dilaton).

This action is obviously invariant under conformal transformations when we 
take the matter fields to transform in the standard way under {\it general}
coordinate transformations, treating conformal transformations as a 
subgroup. In fact the action is invariant under general coordinate 
transformations.
However, one might think  that there are other conformally 
invariant terms possible which cannot be written as general coordinate 
invariants. But this is not the case to two-derivative order \cite{ellis}. 
One, more or less brute force, way to see this is to  
 systematically list all terms of this order subject only 
to rigid scale and Poincare invariance. All terms which are 
independent of those in Eq. (\ref{full}) then contain derivatives of $\phi$, 
are 
straightforwardly seen to be 
non-invariant under special conformal transformations and are 
therefore excluded. Thus full general coordinate invariance is an 
accidental (gauge) symmetry of matter couplings at two-derivative order. 
In fact I suspect this remains true to all orders, but there is as yet no
general proof of this. 
For our purposes it will be sufficient to work to two-derivative order and 
treat higher orders as ``beyond experimental precision''.  

Accidental general coordinate invariance for the matter couplings to the 
dilaton clearly translates into an accidental Equivalence 
Principle for light matter 
emerging from the broken conformal theory. 
 That is, light matter sees the dilaton only 
via the generally covariant couplings to the auxiliary metric $g_{\mu \nu}$.
For a fixed dilaton field, non-linearly realized conformal invariance forces 
the matter fields to propagate and interact ``as if'' 
they were in a curved space with the auxiliary 
metric. 


\section{Newton's Law}

Let us consider first a simple example with 
a single scalar species of matter, $\chi$. Eq. (\ref{full}) clearly 
allows matter to 
have mass (since conformal invariance is spontaneously broken).
Suppose there are several
non-relativistic $\chi$ particles in some reference frame, 
widely-separated  so that we can 
neglect $\chi$ 
self-interactions. We also take $\Lambda$ (now thought of as the 
VEV of $V(\chi)$) to be negligibly small. 
Then Eq. (\ref{full}) can be written as 
\begin{equation}
{\cal L}_{eff} = \frac{\phi^2}{2} (\partial \chi)^2 - \frac{m^2}{2} \phi^4 
\chi^2 + \frac{M^2}{2} (\partial \phi)^2 .
\end{equation}
Doing the field redefinition $\chi \phi \rightarrow \chi$ and some integration
by parts, 
\begin{equation}
{\cal L}_{eff} = \frac{1}{2} (\partial \chi)^2 - \frac{m^2 \phi^2}{2} 
\chi^2 + \chi^2 \frac{\partial^2 \phi}{2 \phi} + 
 \frac{M^2}{2} (\partial \phi)^2.
\end{equation}
This contains trilinear vertices $ - m^2 \pi(x) \chi^2/M + \chi^2 \partial^2 
\pi/2M$ from which we can construct one-dilaton exchange diagrams between 
a pair of $\chi$ particles. Clearly such exchanges will be ultra-local if 
we use the second vertex, so we use just the
first.  After amputating external $\chi$ legs we get 
$m^2/(M^2 \vec{q}^2)$, corresponding to a non-relativistic Newtonian 
potential in position space, $m^2/(M^2 r)$, and confirming our earlier claim 
that $M$ is the analog ``Planck scale''.

This result is much more general. There may be several types of
 non-relativistic particles, $\Psi$, of mass $m$ which
 can carry 
spin and may even 
be composites of the fundamental fields $\chi, \psi, A_{\mu}$. We can write 
a heavy particle   effective theory \cite{hpet}, constrained only by 
the fact that all matter sees the 
dilaton via  generally covariant couplings to  
$g_{\mu \nu} \equiv \phi^2 \eta_{\mu \nu}$, 
\begin{equation}
{\cal L}_{eff} = \sqrt{-g} \{ \frac{g^{\mu \nu}}{2m} (\partial_{\mu} - i m 
v_{\mu}) \Psi^{\dagger}_{v} (\partial_{\mu} + i m 
v_{\mu}) \Psi_v - \frac{m}{2} \Psi^{\dagger}_{v} \Psi_v 
+ {\rm less ~ relevant} \}, 
\end{equation}
where $v_{\mu}$ is a four-velocity defining a frame in which the $\Psi$ 
are non-relativistic. Such contributions for different species of $\Psi$ are 
implicitly taken to be summed here. 
Spin degrees of freedom decouple at leading order. 
A derivation 
of such effective lagrangians in the general relativistic context is 
given in Ref. \cite{fat}, based on earlier discussions in \cite{hpetgr} 
\cite{97}. After field redefining $\Psi_v \rightarrow 
\Psi_v/\phi$, the canonical (heavy particle) fields have a leading 
trilinear vertex, $ m \pi(x) |\Psi_v|^2/M$, which results 
again in the general form of Newton's Law, $m m'/(M^2 r)$.

Of course, the Newtonian approximation displays the non-relativistic face of 
the Equivalence Principle. Even though relativistic effects distinguish 
the dilaton theory from real gravity, the Equivalence Principle remains true 
in the relativistic regime in both theories by (accidental) 
general covariance.

There is one important difference between scalar and real gravity when we 
cannot neglect the gravitational self-energy of the non-relativistic 
``particles'', for example when these ``particles'' are  whole planets 
or stars. In real gravity our effective Lagrangian continues to hold 
because full general covariance is exact. However, for scalar gravity the 
derivation of the effective Lagrangian 
only holds when the auxiliary metric is treated as a 
background field. (Of course once derived we can use it to derive Newton's 
Law by integrating out $\phi$.) Therefore it does not hold when 
gravitational self-energy is significant. 
Still, these self-energies are typically small and 
 many tests of the Equivalence 
Principle (say rates of fall
 of modest masses in the Earth's gravitational field) 
 are experimentally insensitive to this difference, so the analogy is 
good. We will explicitly see in the next section that there is a 
 (small) inequivalence of gravitational and 
inertial masses when gravitational self-energy is taken into account, in 
contrast to real General Relativity.

\section{At the Schwarzchild Radius}

For simplicity, let us determine the spherically symmetric dilaton field 
about a point mass. 
While quantum matter fields ultimately do describe point particles, 
the most direct approach to the classical regime is to employ the 
classical point-particle action functional of the particle worldline 
$x^{\mu}(\tau)$, 
\begin{eqnarray}
S_{particle} &=& 
- m \int d \tau \sqrt{g_{\mu \nu}(x(\tau))~ \frac{dx^{\mu}}{d \tau}
\frac{dx^{\nu}}{d \tau}} \nonumber \\
&=& - m \int d\tau~ \phi(x(\tau)) \sqrt{\eta_{\mu \nu} \frac{dx^{\mu}}{d \tau}
\frac{dx^{\nu}}{d \tau}}.
\end{eqnarray}
In the absence of $\Lambda$ and in the static
limit for the particle-dilaton 
system, 
$t(\tau) = \tau, \vec{x}(\tau) = \vec{0}$, $\phi = \phi(\vec{x})$, 
the dilaton equation of motion reads,
\begin{equation}
\nabla^2 \phi = \frac{m}{M^2} \delta^3({\vec{x}}). 
\end{equation}
The spherically symmetric solution in polar coordinates is then
\begin{equation}
\phi(r) = 1 - \frac{m}{4 \pi M^2 r},
\end{equation}
where the behavior at infinity is chosen to match onto the vacuum $\phi = 1$.
Note that this is an exact solution of the equations of motion to leading 
order in the derivative expansion.

Now let us try to compute the mass of the particle including its dilaton 
field and see if it matches the ``gravitational mass'' setting the 
coefficient of the $1/r$ fall off, which is clearly just $m$. The Hamiltonian
of our theory is given (after implementing the static ansatz) by
\begin{eqnarray}
H &=& \int d^3 \vec{x} \{ \frac{M^2}{2} (\nabla \phi)^2 + m \phi 
\delta^3(\vec{x}) \} \nonumber \\
&=& \int d^3 \vec{x} \{ - \frac{M^2}{2} \phi \nabla^2 \phi + m \phi 
\delta^3(\vec{x}) \} + \frac{M^2}{2} \int^{\infty} d^2 \vec{S}. 
\phi \nabla \phi, 
\end{eqnarray}
where we have performed an integration by parts in the second equality and 
the last term is the surface term at infinity. Now let us 
plug in our dilaton solution to get the rest-energy or mass, 
using the equation of motion to simplify 
the computation of the volume term, 
\begin{equation}
E = \frac{m}{2} (1 + \phi(0)). 
\end{equation}
This is ill-defined because $\phi$ diverges at the origin. If we 
consider our calculation to be only an approximation to a finite-sized 
but compact mass, then the divergence is cut off 
by some length scale, $r_m$,   
\begin{equation} 
E = m(1 - \frac{m}{8 \pi M^2 r_m}).
\end{equation}
This exhibits a non-vanishing gravitational correction relative to the 
gravitational mass, $m$. By contrast in 
 ordinary General Relativity the gravitational mass setting
 the $1/r$ fall-off is 
exactly the same as rest-energy of the system.

There is a second type of singularity in our solution which is more 
alarming. Our dilaton solution passes through zero at a finite 
distance from the mass, of order the usual Schwarzchild radius, 
\begin{equation}
r_{S} = \frac{m}{4 \pi M^2}.
\end{equation}
Since the local scale of spontaneous breaking of conformal invariance is 
set by $\phi(x) M$, whose non-vanishing justifies our chiral Lagrangian 
description, we really cannot trust our solution near or inside 
$r_{S}$. This is in contrast to General Relativity where curvatures
are low at the Schwarzchild radius and the Schwarzchild solution can 
still be trusted. However, the two types of gravity 
are still somewhat analogous in that something very interesting happens 
at the Schwarzchild radius in each case when one compresses mass within 
or near this radius. 

Of course the simple way to avoid this singularity is to 
consider a mass of finite size larger than $r_{S}$, so that the 
interior solution is modified and all singularities smoothed out, as for 
example would be the case for any conventional star or planet. I do not 
know how to derive any exact solutions for finite size objects with some 
reasonable equation of state, although one can work perturbatively. 
But one might also wonder what happens 
if matter collapses inside $r_S$. There appears to be no 
robust answer without reference to the details of the conformal theory in 
its symmetric phase, which is being restored as $\phi \rightarrow 0$ near 
$r_S$.
 One expects the fundamental conformal theory 
beyond the chiral Lagrangian description to become important near 
$r_{S}$ and to resolve all singularities. In Section 15 we will see 
some hints that the nature of this resolution is to replace 
the singularities by something like a black hole, but not black holes 
derivable purely within the chiral Lagrangian (which obviously do not exist).

Since the dilatonic effective theory does not 
possess black holes, the usual arguments in favor of the 
Holographic Principle \cite{holo} do not apply. And yet, as we have seen, 
the effective field theory breaks down if one tries to pack mass inside 
its ``Schwarzchild radius'', $r_S$. That is, most of the naive states of 
the effective field theory are in fact not sensibly described by it and 
the full CFT must take over. 
This reduction of the degrees of freedom within effective 
theory control would be 
interesting to study more precisely.

\section{Light Unbent} 

Our dilaton theory is fundamentally relativistic and has the same 
non-relativistic limit as ordinary gravity. However, of course, the 
relativistic details differ, unlike real gravity there is only scalar 
``gravitational'' radiation here. Perhaps more significantly, light does 
not bend in a dilaton field, where ``light'' can refer to any free (at low 
energies) massless  vector field surviving conformal symmetry breaking in the 
CFT. In fact there is no interaction with the dilaton at all.
This is easy to see, and well known as the Weyl invariance of the 
minimally-coupled Maxwell action: 
\begin{eqnarray}
{\cal L}_{light} &=& 
- \sqrt{-g} \frac{g^{\mu \alpha} 
g^{\nu \beta}}{4} F_{\mu \nu} F_{\alpha \beta} \nonumber \\
&=& - \frac{\eta^{\mu \alpha} \eta^{\nu \beta}}{4}
 F_{\mu \nu} F_{\alpha \beta}. 
\end{eqnarray}

Of course, the dilaton will couple to light in higher dimension 
effective operators such as
\begin{eqnarray}
{\cal L}_{higher-dim.} &\propto& \sqrt{-g} R g^{\mu \alpha} 
g^{\nu \beta}  F_{\mu \nu}  
F_{\alpha \beta}, 
\end{eqnarray}
but the effect rapidly becomes negligible for dilaton fields
softer than the scale suppressing the higher dimension operator.
However, there are circumstances where such effects could 
be magnified. For example in the dilaton field due to a star not much 
larger than $r_{S}$, $\phi$ becomes small and  the 
higher derivative operator becomes important because it scales like 
$1/\phi^2$. The qualitative
difference with General Relativity is that this bending cannot be 
predicted quantitatively within the effective theory since it arises 
from  non-minimal higher dimension operators. Again, in 
Section 15 we will see that with some more information about the 
fundamental conformal dynamics comes a greater level of predictivity.



\section{Quantum Effective Field Theory}

The dilaton couplings are certainly non-renormalizable, but just like 
General Relativity coupled to matter, they can be treated by the standard 
methods of quantum effective field theory in the sub-Planckian regime 
\cite{dono}. The 
consistency of the (non-linearly realized) 
conformal invariance of the effective theory in the 
quantum regime is easy to prove. Using our Goldstone field, the dilaton, 
dimensional regularization can be made fully covariant. The simplest 
way to see this is to work in the auxiliary metric language and dimensionally 
regulate in the manner familiar in General Relativity, 
\begin{eqnarray}
{\cal L}_{(4+\epsilon)D} &=& \mu^{\epsilon} \sqrt{-g} \{ - \frac{M^2}{12} R + 
k_1(\chi, \psi) g^{\mu \nu} D_{\mu} \chi^*  D_{\nu} \chi +
k_2(\chi, \psi) \overline{\psi} i e^{\mu}_{~ a} D_\mu \gamma^a \psi   
- V(\chi, \psi) \nonumber \\
&-& k_3(\chi, \psi) g^{\mu \alpha} 
g^{\nu \beta} F_{\mu \nu} F_{\alpha \beta} 
- k_5(\chi, \psi) \overline{\psi} \sigma^{ab} \psi F_{\mu \nu}  e^{\mu}_{~ a}
e^{\nu}_{~ b} \nonumber \\
&+& k_6(\chi, \psi) e^{\mu}_{~ a} D_{\mu} \chi \overline{\psi} \gamma^{a} \psi
+ ~ {\rm higher ~ derivatives} \},
\end{eqnarray}
where $g_{\mu \nu} \equiv \phi^2(x) \eta_{\mu \nu}$ now is a $(4+\epsilon) 
\times (4+\epsilon)$ matrix, $\mu$ is the RG scale, 
and we have dropped the $k_4$ term of Eq. (\ref{full}) in order 
to avoid the usual issue to do with dimensionally continuing 
the $\epsilon$-symbol.
The central impact of this is that now $\sqrt{-g} = \phi^{4 + \epsilon}$, 
the extra $\phi^{\epsilon}$ multiples $\mu^{\epsilon}$, effectively 
turning the latter
 into a spacetime dependent RG scale! Thus the naive lack of 
conformal invariance in general matter couplings arising from the usual 
scale anomaly, and tracked by $\mu$-dependence, is precisely cancelled 
by the accompanying $\phi$-dependence. A related discussion in a 
different context appears in Ref. \cite{tomboulis}. 
The way in which all this works out 
after renormalization is simply illustrated by considering a matter sector 
consisting of QED, to which we now turn.

\subsection{Anomaly-mediated bending of light}

We will consider here quantum matter in the form of QED, coupled to a 
soft
dilaton background. We can neglect the quantum 
dilaton backreaction if we consider 
QED processes far below $M$. For a constant $\phi$ background
the (renormalized) 
vacuum polarization is easily seen to be the same
 as in pure QED but with the rescalings 
$m_{electron} \rightarrow m_{electron} \phi, \mu \rightarrow \mu \phi$, 
\begin{equation}
\Pi_{\mu \nu}(q) = 
\frac{e^2(\mu)}{2\pi^2} (\eta_{\mu \nu} q^2 - q_{\mu} q_{\nu})
 \int_0^1 dx x(1-x) \ln (\frac{m^2 \phi^2 - q^2 x(1-x)}{\mu^2 \phi^2}).
\end{equation}


First consider the case $m_{electron} = 0$. 
Then in position space the renormalized effective action (following from 
the above vacuum polarization) 
is given by
\begin{equation}
\Gamma = \int d^4 x \frac{e^2(\mu)}{6 \pi^2} \ln (\mu \phi)
F_{\mu \nu}(x) F^{\mu \nu}(x)  + \mu \phi-{\rm independent}.
\end{equation}
(Note, the $\mu \phi$-independent terms contain finite non-local 
pieces as well as the classical action.)
This remains true even when $\phi(x)$ is slowly varying in spacetime, 
\begin{equation}
\Gamma =  \frac{e^2(\mu)}{6 \pi^2} \ln (\mu \phi(x))
F_{\mu \nu}(x) F^{\mu \nu}(x)  + \mu \phi-{\rm independent},
\end{equation}
the locality of the $\phi$-dependent terms guaranteed by the locality of the
ultraviolet divergences that necessitate the accompanying $\mu$-dependence. 
Note that although the effective action is non-analytic in $\phi$, as long 
as we are expanding around a vacuum $\langle \phi \rangle \neq 0$, we can 
expand in a power series in the canonical $\pi(x)$ field, that is effective
vertices involving $\pi$. The quantum effects of massless charges cause
 light to bend in a
dilaton field!


We see here that the interacting photon does couple to the dilaton in
order to cancel the  
conformal anomaly in pure QED, and we can therefore use the RG functions 
which describe this anomaly to determine the dilaton coupling. 
We therefore say that this sensitivity to the dilaton is ``anomaly-mediated''.
There 
is a closely related phenomenon in standard supergravity, 
``anomaly-mediated supersymmetry breaking'' \cite{amsb}, where instead of tracking 
matter sensitivity to the  
$x$-dependence of the dilaton background we track matter sensitivity to 
 the supergravity background dependence on superspace Grassman coordinates. 
The role of the dilaton is played by 
the auxiliary chiral supermultimplet of supergravity known as the 
``compensator'' \cite{compensate}.


Now consider the case $m_{electron} \neq 0$. Then for soft light 
the $\phi$-dependence 
cancels out of the vacuum polarization and the photon is decoupled 
from the dilaton even at the one-loop level. 
 The higher dimension operators such as that in Section 6 can be induced 
also by loops upon integrating out the electron. We have neglected 
such effects in computing the polarization by treating $\phi$ as ``nearly'' 
constant.


\subsection{Quantum dilaton effects}

There is no obstacle to including dilaton internal lines in Feynman diagrams. 
Dimensional regularization continues to provide a conformally invariant 
regularization  when implemented as described above. Just as for standard 
quantum gravity corrections to Feynman diagrams, dilaton exchanges are 
Planck-suppressed and therefore irrelevant in the far infrared. It is only 
when considering such effects that the non-renormalizability of the effective 
theory becomes an essential complication and the theory must be treated 
by the standard methods of non-renormalizable effective field theory. But 
there are no extra subtleties compared with the 
treatment of quantum gravity \cite{dono} in this regard, in fact there are 
fewer.

\section{Symmetric Spacetimes and Cosmological Horizons}

In previous sections we have mostly neglected the non-derivative dilaton
coupling, $\Lambda$. Let us now consider simple solutions in its 
presence. 
In the absence of matter (or after integrating out matter effects), the 
dilaton equations of motion are 
\begin{equation}
\partial^2 \phi = - \frac{4 \Lambda}{M^2} \phi^3.
\end{equation}
Clearly if $\Lambda \neq 0$ there are no Poincare invariant solutions with 
spontaneously broken conformal invariance, $\phi \neq 0$. This is 
analogous to the absence of Poincare invariant solutions in ordinary gravity 
when the cosmological constant is non-zero. 
However, there are simple 
solutions without Poincare invariance, 
\begin{eqnarray}
\phi &=& \frac{M}{\sqrt{- 2 \Lambda} t}, ~ \Lambda < 0 \\
\phi &=& \frac{M}{\sqrt{2 \Lambda} z}, ~ \Lambda > 0
\end{eqnarray}
where $t$ (or  $-t$) is Minkowski time, 
and $z$ is a Minkowski spatial coordinate. 
When  matter is part of the effective theory 
 it only sees the auxiliary metrics, 
\begin{eqnarray}
g_{\mu \nu} &=& - \frac{M^2}{2 \Lambda t^2} \eta_{\mu \nu},~ \Lambda < 0 \\
g_{\mu \nu} &=& \frac{M^2}{2 \Lambda z^2} \eta_{\mu \nu},~ \Lambda > 0, 
\end{eqnarray}
which are (patches of) the maximally symmetric 
spacetimes, $dS_4$ and $AdS_4$ respectively.

Clearly $\Lambda$ behaves quite analogously to the cosmological constant in 
ordinary gravity, and from now on we will refer to it as such. One 
difference to note is that the $dS_4$ is associated here with $\Lambda < 0$
and $AdS_4$ with $\Lambda > 0$, the opposite of the familiar correlation in 
gravity. The reason is simply traced to the fact that when written in terms of
the auxiliary metric the dilaton kinetic term has the opposite sign from 
the usual Einstein action for the reasons discussed at the end of Section 2.

Obsevers made out of light matter in our $dS_4$ auxiliary spacetime will 
see the usual cosmological horizon and quantum mechanically will see 
the associated Hawking radiation \cite{hawking} since this phenomenon is only 
a consequence of doing matter field theory in the background geometry. 
They will infer a finite
 entropy and wonder how to microscopically account for it.

\section{Cosmologies}

If $\Lambda$ is truly a constant then the $dS_4$ solution describes a state 
of permanent inflation. However, what is usually meant by inflation 
occurs when a metastable matter vacuum dominates the energy density, 
but ends when this state relaxes in some way to a true vacuum with 
negligible vacuumm energy. Clearly this is impossible in the dilaton theory 
because inflation is caused by negative energy density which cannot relax to 
zero.

Let us seek other homogeneous and isotropic cosmological solutions 
in (non-relativistic) matter-dominated and radiation-dominated regimes, 
and compare them to standard (spatially flat say) 
Friedman-Robertson-Walker cosmologies. Cleary $\phi$ will be a function of 
time only.
To facilitate the comparison note that 
since 
matter and radiation couple to the dilaton via their generally covariant 
couplings to the auxiliary metric, they cannot distinguish this metric from 
a  non-conformal time-coordinate transformation of the metric, 
\begin{equation}
ds^2 = d\tau^2 - a^2(\tau) d \vec{x}^2, 
\end{equation}
where $d \tau/dt = \phi(t), ~ a(\tau) = \phi(t)$.  
In a matter-dominated era we have the usual scaling of the energy density, 
$\rho(\tau). a^3(\tau) = \rho_0 =$ constant. That is, $\rho_0$ is the fixed 
mass density of matter with respect to {\it coordinate-volume}. 
This 
allows us to straightforwardly generalize the point source of Section 5 to 
this density using superpostition, which imposing homogeneity yields, 
\begin{equation}
\partial_t^2 \phi = - \frac{\rho_0}{M^2}. 
\end{equation}
The simple solution is 
\begin{equation}
\phi(t) = - \frac{\rho_0}{2 M^2} t^2 + \phi_1 t + \phi_0, 
\end{equation}
where $\phi_{0,1}$ are integration constants.
By a shift of the origin of the time coordinate we can put this in the 
simpler form,
\begin{equation}
\phi(t) = \phi_0 - \frac{\rho_0}{2 M^2} t^2. 
\end{equation}

In standard gravity, matter domination gives $a \propto \tau^{2/3}$, which 
is easily seen to correspond to $\phi \propto t^2$. This appears
 similar to the scalar gravity result but there is a crucial sign difference 
in the $t^2$-dependence, once again traced to the fact that the dilaton has 
canonical-sign kinetic term while the conformal mode of standard gravity does 
not. In the scalar gravity case we see that the physical regime, $\phi > 0$, 
gives the universe a finite lifetime between a Big Bang and a Big Crunch, 
where we have $\phi = 0$ and can therefore not trust the chiral Lagrangian
 predictions. But these cosmological 
singularities must be somehow resolved by the fundamental 
conformal field theory dynamics.

Let us now consider radiation dominance. This case is even easier, because 
as we have earlier noted, radiation does not couple to the dilaton. Therefore
we have simply, 
\begin{equation}
\partial_t^2 \phi = 0,
\end{equation}
with solution 
\begin{equation}
\phi = \phi_1 t + \phi_0.
\end{equation}
If $\phi_1 \neq 0$ then we can remove $\phi_0$ by  a shift of the time 
coordinate. This clearly yields 
\begin{equation}
a(\tau) = \sqrt{2 \phi_1} \tau^{1/2}, 
\end{equation}
which is very similar to the standard 
radiation-dominated FRW cosmology, but the 
Hubble parameter  is not related to the radiation energy density. 
There is another possible solution when $\phi_1 = 0$, which yields
\begin{equation}
a(\tau) = \phi_0,
\end{equation}
which corresponds to Minkowski space, despite the presence of radiation.

As will be discussed in Section 15, there is a higher-dimensional 
embedding of the four-dimensional dilaton effective theory, in which form the 
above cosmological solutions were first studied \cite{csaba}. 
In particular, it was pointed out that the crucial sign differences  we 
have seen occurring in  
the scalar and standard cosmologies implies a very different unfolding of 
the universe and condensation of its elements.

\section{The Cosmological Constant Problem}

As discussed in Section 2, taking $\Lambda \ll M^4$ as we must to have a 
sensible effective field theory of the broken phase of a CFT, appears 
unnatural. With the inclusion of matter we see that we are dealing with 
an almost exact analog of the usual cosmological constant problem. Matter 
couples to the dilaton via the auxiliary metric $g_{\mu \nu} \equiv 
\phi^2 \eta_{\mu \nu}$ in exactly the usual generally covariant fashion, 
so the matter vacuum energy, classical plus quantum 
corrections, contributes to our cosmological constant 
$\Lambda$ exactly as it would in the case of the same matter coupled to  
real gravity. An important qualification is that  
the equality of matter vacuum energy contributions to 
the dilatonic and standard gravity cosmological constants is only 
guaranteed if one uses the same UV regularization, for example 
dimensional regularization. Not only are the contributions 
the same, the effects of these constants in 
obstructing Poincare invariant solutions to the equations of motion is also 
very similar, as
 discussed in the previous section. Of course, the
 subleading (Planck-suppressed) 
quantum gravity and quantum dilaton corrections 
to the matter vacuum energy are certainly different in detail, but this is 
not the most robust aspect of the cosmological constant problem.

I find this close analogy of the technical face of the cosmological constant 
problem very  tantalizing. In a sense the analog of gravity is now 
much simpler, just a scalar field. 
Yet the cosmological constant problem seems 
essentially the same, and just as hard. Is it the case nevertheless 
that other
differences between real gravity and the dilaton are essential for 
solving the problem in the former case, or can both the real and analog 
problems be solved by the same basic mechanism?

Ref. \cite{weinberg} derived 
a useful No-Go theorem to filter out a large class of 
proposals for dynamical adjustment of the cosmological constant 
 by light matter fields. The derivation makes central use of the trace of 
Einstein's equations. This is precisely the Einstein equation which 
follows by varying metrics of the special form $g_{\mu \nu} = 
\phi^2(x) \eta_{\mu \nu}$. Further, the derivation is insensitive to the 
sign flip discussed in Section 2 between the dilatonic and standard Einstein 
kinetic terms. Therefore the derivation and no-go theorem apply to 
the dilatonic theory.

\section{Ultraviolet Completion}

Our non-renormalizable effective theory of the dilaton coupled to other light 
remnants of spontaneous conformal invariance breaking must be UV-completed 
by a conformal field theory (CFT) with a continuum limit. To have a broken 
phase it must possess a (at least approximate) 
moduli space of degenerate vacua parametrized by the VEV of the corresponding 
Goldstone boson, $\phi$. The moduli space may have several such directions 
breaking conformal invariance. At the origin of moduli space the conformal 
invariance is intact, but anywhere else it is spontaneously broken.

Such a CFT would be
 to scalar  gravity what string theory is 
for real gravity, the UV completion of the non-renormalizable  
effective
theory.
Just as with string theory, the simplest constructions come with 
supersymmetry, for example $N=4$ super-Yang-Mills theory or the 
conformal field theories arising in $N=1$ SQCD in the conformal window. It 
is much more difficult to find interacting, fundamentally non-supersymmetric
 CFT's with (approximate) moduli spaces. Just as in the 
case of string theory and real gravity, 
this does not  prove that such non-supersymmetric 
theories do not exist, but certainly the supersymmetric examples are easier to
find. 

At first sight it may seem like a simplification that the UV completion of 
the scalar gravity theory can still be a field theory, just a CFT, while 
quantum gravity requires going outside field theory to string theory. 
However, the distinction has somewhat diminished 
with the advent of the AdS/CFT correspondence \cite{adscft}. See Sections 
15 and 16.

\section{Sub-Planckian length scales?}

In ordinary quantum gravity it is difficult to attach meaning to 
distances smaller than the Planck length. 
In the dilaton
 theory, at distances smaller than $1/M$ the non-renormalizable chiral 
lagrangian description certainly breaks down,
but the fundamental CFT description is still valid 
and there is a  Minkowski spacetime in which CFT matter propagates. 
But even though the theory is under control,
 the restored conformal invariance at distances below $1/M$
 certainly makes distance a less meaningful experimental quantity, whereas 
at larger distances the concept is as useful as in the real world. 
Earlier studies of the distance limit in real gravity \cite{pad}, 
but restricted to 
the quantum dynamics of the conformal factor for simplicity, 
naturally relate to the case of the dilaton theory.

\section{Dilaton Mass}

Recently it has been shown \cite{gravmass} that  one can make effective field theory sense
of quantum General Relativity weakly 
deformed 
by  graviton interactions that violate general coordinate invariance. In 
particular, the graviton can be given a small mass. One price for this 
violation of the ``gauge symmetry'' is that the cutoff imposed by 
non-renormalizability on the effective theory is lowered below the 
Planck scale to a weighted geometrical mean of the Planck scale and the 
graviton mass. A second price is that although string theory gives a good 
account of what might UV complete ordinary General Relativity, there is no 
known candidate for such a completion in the deformed case. 

The analogous issues in the dilaton case are clearer and more satisfying, 
and perhaps may shed light on how things might ultimately work in real 
gravity.
The basic plot is that the fundamental UV theory is not a CFT but rather 
an asymptotically free theory which flows to an
infrared attractive CFT. This CFT contains a moduli space as discussed in the
last section. We imagine living away from the origin of moduli space in a 
field direction which we call $\phi$, the dilaton.
 An explicit 
example of such a situation is given by SQCD theories in the
conformal window. Thus, the protective symmetry of our dilaton effective 
field theory, namely (spontaneously broken) conformal invariance, is not an 
exact symmetry of the dynamics, but rather an accidental IR symmetry. 
If one considered the vacuum at the origin of the moduli space then the RG flow of the 
fundamental UV theory would get arbitrarily close to the 
attractive fixed point 
associated with the CFT. However, with $\langle \phi \rangle \neq 0$, this 
RG flow is interrupted at $\langle \phi \rangle M$, so that there is some residual 
deviations from exact conformal invariance of the dynamics. These 
residual deviations can lead to a stabilized dilaton with non-zero mass 
\cite{nima} \cite{rattazzi}. 

In more detail \cite{rattazzi}, suppose the fundamental UV theory gets 
 close enough to the 
infrared fixed point once we have run down to a scale $\Lambda_{CFT}$
that we can begin 
trusting the RG flow 
of the CFT linearized about the associated fixed point. We can write the 
effective lagrangian at this scale as  
\begin{equation}
{\cal L}(\Lambda_{CFT}) = {\cal L}_{fixed-point} + \sum_n g_n(\Lambda_{CFT}) {\cal O}_n(x),
\end{equation}
where the ${\cal O}_n(x)$ are a basis of scaling operators for the CFT, which 
are irrelevant since the CFT is IR-attractive. The dimensionless 
coefficients $g_n(\Lambda_{CFT}) \sim {\cal O }(1)$  
(so that we are at the border 
of being able to linearize the RG in these couplings). This linearized flow is
given by 
\begin{equation}
\mu \frac{d}{d \mu} g_n(\mu) = \gamma_n g_n(\mu), 
\end{equation}
where $\gamma_n > 0$ are the anomalous dimensions (deviations of the scaling 
dimensions from four) of the associated operators. They govern the running 
down to the scale $\langle \phi \rangle  M$  where the effective 
chiral lagrangian for the dilaton and other light remnants takes over. Running
down to $\langle \phi \rangle M \ll \Lambda_{CFT}$ yields 
\begin{eqnarray}
{\cal L}(\langle \phi \rangle M) &=& {\cal L}_{fixed-point} + 
\sum_n g_n(\langle \phi \rangle M) {\cal O}_n(x)
\nonumber \\
&=& {\cal L}_{fixed-point} + \sum_n (\frac{\langle \phi \rangle 
M}{\Lambda_{CFT}})^{\gamma_n}
 g_n(\Lambda_{CFT}) {\cal O}_n(x).
\end{eqnarray}

To go below the $\langle \phi \rangle M$ threshold we must match onto the dilatonic effective 
theory but now including the small perturbations with different scaling 
properties from the exact CFT. For simplicity let us consider only the 
most relevant of the operators ${\cal O}_n$, calling it simply $\hat{\cal O}$, 
with coupling $g$ and anomalous dimension $\gamma > 0$. This perturbation 
can and generically will match onto  a small 
correction to the dilatonic effective theory, 
the most relevant such effect being on the dilaton potential, 
\begin{equation}
{\cal L}_{dilaton} = \frac{M^2}{2}(\partial \phi)^2 - \Lambda \phi^4 + 
{\cal O}(1) (\frac{\phi M}{\Lambda_{CFT}})^{\gamma} g(\Lambda_{CFT}) 
M^4 \phi^{4} 
+ {\rm less ~ relevant},
\end{equation}
where we have used the fact that $M \phi$ is the only scale (up to 
derivatives) that can  
saturate the canonical dimension (four) of $\hat{\cal O}$ in the 
matching to the effective Lagrangian.
It is straightforward to see that the resulting potential violates conformal 
invariance and can stabilize the dilaton 
in the broken phase and generate a small mass for it \cite{rattazzi} 
(if $\Lambda$ is small 
as we have assumed all along).

\section{Supersymmetry (Breaking)}

Making our chiral lagrangian exactly 
supersymmetric is straightforward enough, by 
elevating the dilaton to a chiral superfield. Similarly matter comes in 
supermultiplets. So for example, the pure dilatonic theory would be a massless 
Wess-Zumino theory now, the cosmological constant arising from the 
superpotential. 
The dilaton superfield now couples to matter in a manner 
almost identical to the auxilary conformal compensator of supergravity, 
except that the dilaton multiplet 
has the right-sign kinetic term since it has propagating 
degrees of freedom,  the dilaton, ``axion'' and  dilatino. 
As a consequence, supersymmetric anomaly-mediation \cite{amsb} is 
very similar in supergravity and the dilaton theory. Clearly, unbroken 
supersymmetry can protect the analog cosmological constant just as in 
supergravity theories.

Our central problem occurs if we imagine that 
the matter sector is not supersymmetric or at least has broken supersymmetry, 
just as is the case of the Standard Model in the real world. 
Such supersymmetry breaking 
might arise in one of three ways. 

(a) The UV completion is some (as yet 
unknown) 
non-supersymmetric CFT with an (approximate) modulus. Then there is no mystery
as to why its infrared remnants are non-supersymmetric. This is the 
analog of talking about fundamentally non-supersymmetric strings.

(b) The CFT is exactly 
supersymmetric but, as in the previous 
section, the fundamental theory is not this SCFT but rather an asymptotically 
free theory that flows in the infrared to the SCFT. Now suppose that 
this fundamental theory is NOT supersymmetric, that supersymmetry is just 
an accidental symmetry of the CFT. However, as discussed in the previous 
section the RG flow never reaches the fixed point of the CFT because 
the running stops at $\langle \phi \rangle M$, and therefore the accidental 
supersymmetry does not become exact. Rather it appears in the effective 
dilaton theory as a form of weak but explicit supersymmetric breaking, 
accompanying the weak breaking of the conformal symmetry of the dynamics \cite{markus}. 

(c) The CFT is the UV completion and is supersymmetric but the vacuum we 
expand about is not supersymmetric, that is supersymmetry is spontaneously 
broken. What I find intriguing about this case is that it is 
possible to prove \cite{susybreak} the 
analog of a conjecture that has been made 
in superstring theory \cite{banks}, namely that there are no Poincare 
invariant vacua which are not supersymmetric. That is, 
broken supersymmetry must be accompanied by a cosmological constant which 
obstructs Poincare invariant solutions. 

In the dilaton case a  simple proof can be given 
\cite{susybreak}. Since the fundamental 
SCFT has global
super-Poincare-invariant dynamics (ungauged by supergravity) the order 
parameter for  supersymmetry breaking is simply the (positive) 
vacuum energy density, $\Lambda$. If this does not vanish then it
 also spontaneously breaks conformal invariance. In order to have 
exact (but non-linearly realized) conformal invariance of the dilaton effective lagrangian, this 
vacuum energy must be dressed with the dilaton field to become our analog 
cosmological constant, 
\begin{equation}
{\cal L}_{eff} = \frac{1}{2} (\partial \phi)^2 - \Lambda \phi^4. 
\end{equation}
With such 
a non-zero cosmological constant a Poincare invariant vacuum becomes 
impossible. Of course we can also choose  the ``non-geometric'' phase at the 
origin of moduli space $\langle \phi \rangle = 0$ 
where both supersymmetry and conformal invariance (including Poincare
 invariance) are preserved.


\section{Localized Dilaton in Higher-dimensional Gravity}

The Randall-Sundrum II (RS2) braneworld scenario \cite{rs2} is well known to demonstrate 
how a massless 
4D graviton mode (and attendent 4D General Relativity) 
localized about a 3-brane 
 can emerge from a higher-dimensional gravity theory,  where matter fields 
are taken to be confined to the 3-brane. Macroscopically this leads to 
4D gravity coupled to 4D matter.
What is less well known is that 
there is an exact analog of this where the 4D graviton mode is replaced by 
a 4D dilaton, so that macroscopically we recover our effective dilaton theory.

The localized dilaton model (see Refs. \cite{nima} \cite{rattazzi} for this 
interpretation)
 is reached by beginning with the original RS1 model \cite{rs1} (without 
stabilizing the radius). 
The 4D effective field theory below the masses of KK excitations 
\cite{graesser} \cite{wise} involves
the light matter fields on the ``IR'' brane as well as the massless 
4D graviton and the massless 4D radion, $r_c(x)$, whose VEV is the extra-dimensional
 radius, 
\begin{equation}
{\cal L}_{4D eff} = \sqrt{-g} \{ \frac{1 - e^{-2k \pi r_c(x)}}{k} M_5^3 R 
+ \frac{M_5^3}{k} g^{\mu \nu} \partial_{\mu} e^{-k \pi r_c(x)}
\partial_{\nu} e^{-k \pi r_c(x)} \} + \sqrt{-g_{IR}} 
{\cal L}_{IR matter}(g_{IR}),
\end{equation}
where $g_{IR _{\mu \nu}} = g_{\mu \nu}(x) e^{- 2 k \pi r_c(x)}$ is the induced 
metric on the IR brane and $ {\cal L}_{IR matter}(g_{IR})$ is a 
general  lagrangian for matter localized on the IR brane and covariantly 
coupled to the induced metric. Making the field redefinition
\begin{equation}
\phi \equiv \frac{ e^{-k \pi r_c(x)}}{ e^{-k \pi \langle r_c \rangle }}
\end{equation}
we get, 
\begin{equation}
{\cal L}_{4D eff} = \sqrt{-g} \{ \frac{1 - \phi^2 
e^{-2k \pi \langle r_c \rangle}}{k} M_5^3 R 
+ \frac{M_5^3 e^{-2k \pi \langle r_c \rangle}}{k} g^{\mu \nu} \partial_{\mu} \phi
\partial_{\nu} \phi \} + \sqrt{-g_{IR}} 
{\cal L}_{IR matter}(g_{IR}),
\end{equation}
and 
\begin{equation}
g_{IR _\mu \nu} = g_{\mu \nu}(x) \phi^2.
\end{equation}

Now clearly the effective 4D Planck scale of the 4D graviton is
approximately  $\sqrt{M_5^3/k}$. We can therefore decouple this dynamical 
4D gravity by taking the limit $M_5^3/k \rightarrow \infty$. We will 
choose to do it holding 
$M^2 \equiv 2 e^{-2k \pi \langle r_c \rangle} M_5^3/k$ fixed, and expanding about 
$g_{\mu \nu} = \eta_{\mu \nu}$. Since we have decoupled gravitational 
fluctuations we have exactly $g_{\mu \nu} = \eta_{\mu \nu}$. In this limit 
then,
\begin{equation}
{\cal L}_{4D eff} = M^2
(\partial \phi)^2 + \sqrt{-g_{IR}}~ 
{\cal L}_{IR matter}(g_{IR}),
\end{equation}
and 
\begin{equation}
g_{IR _{\mu \nu}} = \eta_{\mu \nu} \phi^2.
\end{equation}
This is precisely the form of our dilatonic effective field theory.
While in phenomenological RS1 applications one usually chooses the scale we 
defined as $M$ to be of order the weak scale, in our dilatonic analog we 
choose it to be the ``Planck'' scale. The fine-tuning of the analog 
cosmological constant is reflected in the tuning of the IR brane tension in 
RS1, while departures from this tuned case give rise to ``bent brane'' 
behaviors \cite{kaloper} \cite{nihei} that reflect the $AdS_4$ and $dS_4$ 
behaviors we saw earlier. The more general cosmologies we found before 
correspond to the RS1 cosmologies studied in Ref. \cite{csaba} (without 
radius stabilization).

The  limit of RS1 we took, whose 4D effective theory is
 the dilaton effective theory, has a simple spacetime interpretation. We are 
keeping the IR brane fixed and sending the Planck brane infinitely far away.
(It is the opposite of the move we make to get from RS1 to RS2, where we 
keep the Planck brane fixed and send the IR brane infinitely far away.)
Remarkably the radion mode is localized on the IR brane and therefore 
survives this limit and is identified with the dilaton. The localized dilaton
model is this 
limiting case of RS1. The full 5D model also contains massive 
KK gravitons, which of course decouple for the most part 
in the low energy theory. In this 
way there are black holes associated to the higher-dimensional completion of 
the dilatonic theory, 
but they are 5D black holes. Such black holes
 may (I do not know a proof) also  
encompass the IR brane \cite{gidd}, seeded by concentrations of brane matter, 
thereby answering the issue raised in Section 5 of what happens when 
matter lies within its ``Schwarzchild radius'', $r_S$. If this is true 
then the difference with ordinary gravity is in the dimensionality of the 
black hole that forms! Even light, which we saw in Section 6 is usually 
hard to bend, could then be trapped within a 5D black hole encompassing the 
IR brane. In any case in this RS picture it is the KK 
graviton modes  of  the UV completion of the dilaton 
effective theory (in totality a string theory on the RS background)
that  resolves the singularity at $r_S$  (for large $r_S$)
found in Section 5.

\section{AdS/CFT}

We normally approach the AdS $\equiv$ CFT correspondence \cite{adscft} 
starting from 
the left-hand side,  imagining 
string theory as a UV completion 
of quantum gravity in such a background. It is then a surprise that this 
is dual to doing CFT without gravity. The direction is reversed in our 
dilatonic analogy. We imagine doing CFT as the UV completion of our 
dilatonic effective field theory. To our surprise 
(at least if the CFT has a large gap in the spectrum of scaling dimensions and 
a large-$N$ type expansion)
we find that the CFT dynamics is dual to a true gravitational theory 
in AdS$_5$. Since we imagine the CFT to be 
spontaneously broken, the full conformal symmetry of AdS$_5$ is not realized, 
the space is truncated before the horizon. In warped effective field 
theory this is captured as in the previous section \cite{rscft} \cite{nima} 
\cite{rattazzi}.

\section{Euclidean Continuation}

Technically and conceptually it is often useful to 
have a Euclidean continuation of quantum field theory in Minkowski space.
The path integrals associated with the Euclidean continuations are usually 
better defined, allowing some understanding of non-perturbative effects.
In the case of quantum General Relativity the Euclidean path integral 
has a notoriously ill-defined measure for Riemannian metrics, stemming 
from the unboundedness of the Euclidean Einstein action \cite{euclgr}. 
This is due to the ``wrong-sign'' kinetic term for the conformal mode of the
dynamical metric.
There is no such 
problem for the dilaton theory coupled to light 
matter because the dilaton has a canonical kinetic term. 

In Euclidean quantum gravity one expects to sum over different topologies. 
In the dilatonic case, especially in the generally covariant 
formulation given at the end of Section 2, we see that different topologies 
can be related to flat space by globally non-trivial conformal 
transformations. It would be interesting to study whether these non-trivial 
topologies must be included in the dilatonic effective Euclidean path 
integral in order to match the globally non-trivial 
IR consequences of the fundamental CFT.

\section*{Acknowledgements}

I am grateful to Nemanja Kaloper and David E. Kaplan for useful discussions.
This research was supported  by 
NSF Grant P420D3620434350.

\end{document}